\documentstyle[pra,aps,multicol]{revtex}
\renewcommand{\narrowtext}{\begin{multicols}{2} \global\columnwidth20.5pc}
\renewcommand{\widetext}{\end{multicols} \global\columnwidth42.5pc}
\begin{document}
\draft
\title{Probabilistic quantum cloning via Greenberger-Horne-Zeilinger states}
\author{Chuan-Wei Zhang, Chuan-Feng Li\thanks{%
Electronic address: cfli@ustc.edu.cn}, Zi-Yang Wang, and Guang-Can Guo%
\thanks{%
Electronic address: gcguo@ustc.edu.cn}}
\address{Laboratory of Quantum Communication and Quantum\\
Computation and Department of Physics,\\
University of Science and Technology of China,\\
Hefei 230026, People's Republic of China\vspace*{0.3in}}
\maketitle

\begin{abstract}
\baselineskip12ptWe propose a probabilistic quantum cloning scheme using
Greenberger-Horne-Zeilinger states, Bell basis measurements, single-qubit
unitary operations and generalized measurements, all of which are within the
reach of current technology. Compared to another possible scheme via
Tele-CNOT gate $[$D. Gottesman and I. L. Chuang, Nature {\bf 402},{\bf \ }%
390 {\bf (}1999)$]$, the present scheme may be used in experiment to {\it %
clone the states of one particle to those of two different particles} with
higher probability and less GHZ resources.

\end{abstract}

\vskip 1.0cm

\narrowtext
\baselineskip12pt

\section{Introduction}

Quantum computers can solve problems that classical computers can never
solve \cite{Shor94}. However, the practical implementation of such device
need careful consideration of the minimum resource requirement and
feasibility of quantum operation. The basic operation in quantum computer is
unitary evolution, which can be performed using some single-qubit unitary
operations and Controlled-NOT (C-NOT) gates \cite{Bar95}. While single-qubit
unitary operation can be executed easily \cite{Chu95}, the implementation of
C-NOT operation between two particles (for example two photons) encounters
great difficulty in experiment \cite{Mil89}. With linear optical devices
(beam splitters, phase shifters, etc.), the C-NOT operations between the
several quantum qubits (such as location and polarization) of a single
photon is within the reach of current quantum optics technology \cite{Cerf98}%
, but nonlinear interactions are required for the construction of practical
C-NOT gate of two particles \cite{Mil89}. Those nonlinear interactions are
normally very weak, which forecloses the physical implementation of quantum
logic gate.

To solve this problem, Gottesman and Chuang \cite{Got99} suggested that a
generalization of quantum teleportation \cite{Ben93}$-$using single-qubit
operations \cite{Chu95}, Bell-basis measurements \cite{Bra95} and certain
entangled quantum states such as Greenberger-Horne-Zeilinger (GHZ) states 
\cite{Gre90}$-$is sufficient to construct a universal quantum computer and
presented systematic constructions for an infinite class of reliable quantum
gates (including Tele-C-NOT gate). Experimentally, quantum teleportation has
been partially realized \cite{Bou97} and three-photon GHZ entanglement has
been observed \cite{Bou99}. Thus their construction of quantum gates offers
possibilities for relaxing experimental constraints on realizing quantum
computers.

Unfortunately, up to now there has been no way to experimentally distinguish
all four of the Bell states, although some schemes do work for two of the
four required cases $-$yielding at most a $50\%$ absolute efficiency \cite
{Bra95}. In Gottesman and Chuang's scheme, two GHZ states and three
Bell-basis measurements are needed to perform a C-NOT operation, which
yields $1/8$ probability of success in experiment. To complete a unitary
operator, many C-NOT gates may be needed, which makes the probability of
success close to zero. Moreover, the creation efficiency of GHZ states is
still not high in experiment now \cite{Bou99}. Therefore a practical
experiment protocol requires careful consideration of the minimum resource
and the maximum probability of success.

In this paper we investigate the problem of probabilistic quantum cloning
using GHZ states, Bell basis measurements, single-qubit unitary operations
and generalized measurements. The single-qubit generalized measurement can
be performed by the unitary transformation on the composite system of that
qubit and the auxiliary probe with reduction measurement of the probe \cite
{Sch96}. In an optical quantum circuit, the probe qubit can be represented
as the location of a photon and such process can be implemented using only
linear optical components (such as polarizing beam splitter and polarization
rotation) \cite{Cerf98}. We mention above that the construction of practical
C-NOT between two particles is not within current experimental technology,
but it does not prohibit the C-NOT operation between different degrees of
freedom of one photon. This kind of C-NOT is allowed in linear optical
circuit and is of different type from C-NOT between different particles \cite
{Cerf98}. So the single-qubit generalized measurement on the polarization
can be performed with location as the probe.

Consider a sender Alice holds an one-qubit quantum state $\left| \phi
\right\rangle $ and wishes to transmit identical copies to $N$ associates
(Bob, Claire, etc.). Quantum no-cloning theorem \cite{Woot82} implies that
the copies cannot be perfect; but this result does not prohibit cloning
strategies with a limited degree of success. Two most important cloning
machines $-$universal \cite{Buz96,Gis97,Buz97} and state-dependent \cite
{Hil97,Duan98,Che99} $-$have been proposed by some authors. However, it is
not available \footnote{%
However, it is available to clone the states of one qubit of a single photon
to two qubits of that photon using optical simulation \cite{Cerf98}.} for
Alice to generate the copies locally using an appropriate quantum network 
\cite{Buz97,Che99,Zhang00} and then teleport each one to its recipient by
means of teleportation due to the difficulty of executing C-NOT operation 
\cite{Mil89}. To avoid such difficulty, recently, Murao {\it et al.} \cite
{Mur99} presented an optimal $1$ to $N$ universal quantum telecloning
strategy via a $\left( 2N\right) $-particle entangled state. Such
entanglement is difficult to prepare in experiment when $N$ is large.
Quantum probabilistic (state-dependent) cloning machine is designed to
perfectly reproduce linear independent states secretly chosen from a finite
set with no-zero probability \cite{Duan98,Che99,Zhang00}. The corresponding
telecloning process can be executed via the Tele-C-NOT gates \cite{Got99}
according to the cloning strategies provided in \cite{Che99,Zhang00}; but
such procedure requires too many GHZ states and Bell basis measurements and
can succeed with probability close to zero. The scheme we propose in this
paper needs only $\left( N-1\right) $ GHZ states and $\left( N-1\right) $
Bell-basis measurements to implement $M\rightarrow N$ cloning. Although such
process cannot reach the optimal probability as that in local situation, it
may be used in current experiment to {\it cloning the states of one particle
to those of two different particles} with higher probability and less GHZ
resources.

The rest of the paper is organized as follows. In section II we discuss some
strategies of probabilistic cloning and present the concept of {\it %
probability spectrum} to describe different strategies. Comparing two most
important ones, we show that $M$ entries $1\rightarrow N$ cloning give more
copies at the price of higher probability of failure than one $M\rightarrow
N $ cloning. In Section III, we present the probabilistic telecloning
process via three-particle entangled state and also show how to construct
the entangled state from GHZ state by local operations. A summary is given
in Section IV.

\section{Strategies of probabilistic cloning}

Generally, the most useful states are $\left| \phi _{\pm }\left( \theta
\right) \right\rangle =\cos \theta \left| 1\right\rangle \pm \sin \theta
\left| 0\right\rangle $ in quantum information theory. Given $M$ initial
copies, Alice need not to always execute the cloning operation by taking
these copies as a whole. Suppose Alice divides the $M$ copies into $m$
different kinds of shares, each of which includes $\vartheta _i$ entries $%
k_i\rightarrow N_i$ cloning processes. For different kinds of shares, one of
the two parameters $k_i$ and $N_i$ should be different. These parameters
should satisfy 
\begin{equation}
\sum_{i=1}^mk_i\vartheta _i=M\text{.}  \eqnum{2.1}
\end{equation}

The probability of obtaining $x$ copies for Alice can be represented as 
\begin{equation}
P\left( x\right)
=\sum_{\sum\limits_{i=1}^mg_iN_i=x}\prod_{i=1}^mC_{\vartheta _i}^{g_i}\gamma
_{k_iN_i}^{g_i}\left( 1-\gamma _{k_iN_i}\right) ^{\vartheta _i-g_i}, 
\eqnum{2.2}
\end{equation}
where $C_{\vartheta _i}^{g_i}=\vartheta _i!/g_i!(\vartheta _i-g_i)!$, $g_i$
denotes successful cloning attempts in $\vartheta _i$ same processes, $%
\gamma _{k_iN_i}$ is the success probability of $k_i\rightarrow N_i$
cloning, which is 
\begin{equation}
\gamma _{k_iN_i}=\frac{1-\cos ^{k_i}2\theta }{1-\cos ^{N_i}2\theta }. 
\eqnum{2.3}
\end{equation}
$P\left( x\right) $ is the discrete function of $x$ and can be represented
as a series of discrete lines in the $P\left( x\right) -x$ plane, which we
called as {\it Probability Spectrum}. Different probabilistic cloning
strategies are corresponding to different {\it Probability Spectrums}.

Two important parameters can be obtained from {\it Probability Spectrum},
that is, the expected value of the output copies number $E$ and the
probability of failure $F$, which are defined as 
\begin{equation}
E\left\{ k_i,N_i,\vartheta _i\right\}
=\sum_{x=0}^{\sum\limits_{i=1}^m\vartheta _iN_i}xP\left( x\right) , 
\eqnum{2.4}
\end{equation}
\begin{equation}
F\left\{ k_i,N_i,\vartheta _i,K\right\} =\sum_{x=0}^{K-1}P\left( x\right) . 
\eqnum{2.5}
\end{equation}
It is regarded as failure if the copies number Alice attains less than the
cloning goal $K$. When $M$ is large, above two parameters can well describe
different cloning strategies. In the following, we discuss two most
important cloning strategies (the cloning goal $K=N$):

$\left( 1\right) $ cloning the $M$ copies as a whole ($M\rightarrow N$).

$\left( 2\right) $ cloning each copy respectively ($M\times \left(
1\rightarrow N\right) $).

The second is included for it is the strategy we choose in probabilistic
telecloning process. Comparing above two strategies with the two parameters $%
E$ and $F$, we find the second give more copies at the price of higher
probability of failure. In fact, if Alice choose the second strategy, the
cloning attempts may succeed for two or more initial copies, thus Alice may
have chance to get more than $N$ copies. The expected values for the two
different strategies can be represented as 
\begin{equation}
E_1=N\gamma _{MN},  \eqnum{2.6}
\end{equation}
\begin{eqnarray}
E_2 &=&\sum_{k=0}^MkNC_M^k\gamma _{1N}^k\left( 1-\gamma _{1N}\right) ^{M-k} 
\eqnum{2.7} \\
\ &=&MN\gamma _{1N}\sum_{k=1}^MC_{M-1}^{k-1}\gamma _{1N}^{k-1}\left(
1-\gamma _{1N}\right) ^{\left( M-1\right) -\left( k-1\right) }  \nonumber \\
\ &=&MN\gamma _{1N},  \nonumber
\end{eqnarray}
where $2\leq M<N$. Denote $t=\cos 2\theta $, we get $\Delta E=E_2-E_1=N%
\widetilde{\Delta E}/\left( 1-t^N\right) $, where $\widetilde{\Delta E}%
=M-Mt-1+t^M$. Obviously $0\leq t\leq 1$. When $t=0$, $\widetilde{\Delta E}%
=M-1>0$. If $t=1$, $\gamma _{MN}=\frac MN$ and $\Delta E=\widetilde{\Delta E}%
=0$. When $t\neq 1$, $\frac{d\widetilde{\Delta E}}{dt}=-M+Mt^{M-1}<0$, thus $%
\widetilde{\Delta E}$ is monotonously decreasing and always greater than or
equal to zero, that is 
\begin{equation}
E_1\leq E_2,  \eqnum{2.8}
\end{equation}
with equality only for $\left| \varphi _{+}\left( \theta \right)
\right\rangle =\left| \varphi _{-}\left( \theta \right) \right\rangle $ ($%
t=1 $). $\Delta E$ is very large when $M$ is large. The expected values for
different $M$, $N$ are plotted in Fig. 1. 
\[
\text{Fig. 1} 
\]

The failure probabilities of above two strategies are 
\begin{equation}
F_1=1-\gamma _{MN}=t^M\left( 1-t^{N-M}\right) /\left( 1-t^N\right) , 
\eqnum{2.9}
\end{equation}
\begin{equation}
F_2=(1-\gamma _{1N})^M=\left( \left( t-t^N\right) /\left( 1-t^N\right)
\right) ^M,  \eqnum{2.10}
\end{equation}
respectively. Note the fact that for any $a_i\geq 0$, $\left(
\prod_{i=1}^na_i\right) ^{\frac 1M}\leq \frac 1M\left(
\sum_{i=1}^na_i\right) $ with equality only for $a_1=a_2=...=a_n$, we derive 
\begin{eqnarray}
F_1 &=&\frac{t^M}{\left( 1-t^N\right) ^M}\left( 1-t^N\right) ^{M-1}\left(
1-t^{N-M}\right)  \eqnum{2.11} \\
\ &\leq &\frac{t^M}{\left( 1-t^N\right) ^M}\left( 1-\frac{t^{N-M}+(M-1)t^N}M%
\right) ^M  \nonumber \\
\ &\leq &\frac{t^M}{\left( 1-t^N\right) ^M}\left( 1-t^{N-1}\right) ^M=F_2 
\nonumber
\end{eqnarray}
with equality only for $t=0$ or $1$ ($\theta =\pi /4$ or $0$). The failure
probabilities for different $M$, $N$ are illustrated in Fig. 2. 
\[
\text{Fig. 2} 
\]

Now that the two different strategies have both advantage and shortage,
Alice should choose one according to her need. If she need more copies, she
can adopt $1\rightarrow N$ strategy. If she wishes to obtain the copies with
greater success probability, she should choose $M\rightarrow N$ cloning
process.

\section{Probabilistic telecloning process}

Suppose Alice holds $M$ copies of one-qubit quantum state $\left| \phi
\right\rangle _X$ that is secretly chosen from the set $\left\{ \left| \phi
_{\pm }\left( \theta \right) \right\rangle =\cos \theta \left|
1\right\rangle \pm \sin \theta \left| 0\right\rangle \right\} $ and wishes
to clone it to $N$ associates (Bob, Claire, etc.). In local situation, she
can do so using the unitary-reduction operation $-$ a combination of unitary
evolution together with measurements $-$ on the $N+1$ qubit ($N$ qubit of
the cloning system and a probe to determine whether the cloning is
successful) with maximum success probability \cite{Che99} $\gamma
_{MN}=\left( 1-\cos ^M2\theta \right) /\left( 1-\cos ^N2\theta \right) $.
This unitary-reduction operator can be decomposed into the interaction
between two particles using a special unitary gate \cite{Che99}: 
\begin{equation}
D\left( \theta _1,\theta _2\right) \left| \phi _{\pm }\left( \theta
_3\right) \right\rangle \left| 1\right\rangle =\left| \phi _{\pm }\left(
\theta _1\right) \right\rangle \left| \phi _{\pm }\left( \theta _2\right)
\right\rangle .  \eqnum{3.1}
\end{equation}
with $\cos 2\theta _3=\cos 2\theta _1\cos 2\theta _2$ and $0\leq \theta
_j\leq \pi /4$, which suffice to determine $\theta _3$ uniquely. This
operation $D^{\dagger }(\theta _1,\theta _2)$ transforms the information
describing the initial states $\left| \phi _{\pm }\left( \theta _1\right)
\right\rangle \left| \phi _{\pm }\left( \theta _2\right) \right\rangle $
into one qubit $\left| \phi _{\pm }\left( \theta _3\right) \right\rangle $.
With such pairwize interaction, the initial states $\left| \phi _{\pm
}\left( \theta \right) \right\rangle ^{\otimes M}$ can be transferred into
states $\left| \phi _{\pm }\left( \theta _M\right) \right\rangle \left|
0\right\rangle ^{\otimes \left( M-1\right) }$ using corresponding operator $%
D_M=D_1\left( \theta _{M-1},\theta _1\right) D_2\left( \theta _{M-2},\theta
_1\right) ,...D_{M-1}(\theta _1,\theta _1)$, where $D_j\left( \theta
_{M-j},\theta _1\right) $ is denoted as the operator $D\left( \theta
_{M-j},\theta _1\right) $ acts on particles $\left( 1,j+1\right) $ and $%
\theta _j$ is determined by $\cos 2\theta _j=\cos ^j2\theta $. This operator
is unitary and $D_M^{\dagger }$ can perform the reverse transformation. Thus
we only need to transfer the states $\left| \phi _{\pm }\left( \theta
_M\right) \right\rangle $ to the appropriate form $\left| \phi _{\pm }\left(
\theta _N\right) \right\rangle $ to obtain $\left| \phi _{\pm }\left( \theta
\right) \right\rangle ^{\otimes N}$ using operation $D_N^{\dagger }$ (with
similar definition as $D_M^{\dagger }$). This process can be accomplished by
a unitary-reduction operation 
\begin{equation}
U\left| \phi _{\pm }\left( \theta _M\right) \right\rangle _1\left|
P_0\right\rangle =\sqrt{\gamma }\left| \phi _{\pm }\left( \theta _N\right)
\right\rangle _1\left| P_0\right\rangle +\sqrt{1-\gamma }\left|
1\right\rangle _1\left| P_1\right\rangle ,  \eqnum{3.2}
\end{equation}
where $\left| P_0\right\rangle $ and $\left| P_1\right\rangle $ are the
orthogonal bases of the probe system. If a postselective measurement of
probe $P$ results in $\left| P_0\right\rangle $, the transformation is
successful, otherwise the cloning attempt has failed and the result is
discarded. $U$ is unitary and the transformation probability $\gamma =\gamma
_{MN}$. $U$ is a qubit $1$ controlling probe $P$ rotation $R_y\left( 2\omega
\right) =\left( 
\begin{array}{cc}
\cos \omega & \sin \omega \\ 
-\sin \omega & \cos \omega
\end{array}
\right) $ with $\omega =\arccos \sqrt{\frac{\left( 1-\cos ^M2\theta \right)
\left( 1+\cos ^N2\theta \right) }{\left( 1+\cos ^M2\theta \right) \left(
1-\cos ^N2\theta \right) }}$.

Operations $D_M$ and $D_N^{\dagger }$ involve the interactions of two
particles which is difficult to implement in current experiment. In this
paper, we adopt $M\times \left( 1\rightarrow N\right) $ strategy to
substitute $D_M$ and transfer $M$ copies of the states $\left| \phi _{\pm
}\left( \theta \right) \right\rangle $ to $\left| \phi _{\pm }\left( \theta
_N\right) \right\rangle $ respectively using similar unitary-reduction
operation as that in Eq. (3.2). To substitute the operation $D_N^{\dagger }$%
, we use three-particle entanglement to implement the operator $D_j\left(
\theta _{N-j},\theta _1\right) $, which acts as 
\begin{equation}
D_j\left( \theta _{N-j},\theta _1\right) \left| \phi _{\pm }\left( \theta
_{N-j+1}\right) \right\rangle \left| 1\right\rangle =\left| \phi _{\pm
}\left( \theta _{N-j}\right) \right\rangle \left| \phi _{\pm }\left( \theta
_1\right) \right\rangle .  \eqnum{3.4}
\end{equation}

Assume Alice and the $j$-th associate $C_j$ share a three-particle entangled
state $\left| \psi ^j\right\rangle _{SAC_j}$ as a starting resource. This
state must be chosen so that, after Alice performs local Bell measurements
and informs $C_j$ of the results, she and $C_j$ can obtain the state $\left|
\phi _{\pm }\left( \theta _{N-j}\right) \right\rangle _A\left| \phi _{\pm
}\left( \theta _1\right) \right\rangle _{C_j}$ by using only local
operation. Denote $\left| \varphi _i^j\right\rangle =D_j\left( \theta
_{N-j},\theta _1\right) \left| i\right\rangle \left| 1\right\rangle $, $i\in
\left\{ 0,1\right\} $, a choice of $\left| \psi ^j\right\rangle _{SAC_j}$
with these properties may be the three-particle state 
\begin{equation}
\left| \psi ^j\right\rangle _{SAC_j}=\frac 1{\sqrt{2}}\left( \left|
0\right\rangle _S\left| \varphi _1^j\right\rangle _{AC_j}-\left|
1\right\rangle _S\left| \varphi _0^j\right\rangle _{AC_j}\right) , 
\eqnum{3.5}
\end{equation}
where $S$ represents a single qubit held by Alice, which we should refer to
as the ``port'' qubit. The tensor product of $\left| \psi ^j\right\rangle
_{SAC_j}$ with the state $\left| \phi _{\pm }\left( \theta _{N-j+1}\right)
\right\rangle _X=h_j\left| 1\right\rangle \pm t_j\left| 0\right\rangle $ ($%
h_j=\cos \theta _{N-j+1}$, $t_j=\sin \theta _{N-j+1}$) held by Alice is
four-qubit state. Rewriting it in a form that singles out the Bell basis of
qubit $X$ and $S$, we get 
\begin{eqnarray}
&&\left| \Omega ^{\pm j}\right\rangle _{XSAC_j}  \eqnum{3.6} \\
&=&-\frac 12\left| \Psi ^{-}\right\rangle _{XS}\left( h_j\left| \varphi
_1^j\right\rangle _{AC_j}\pm t_j\left| \varphi _0^j\right\rangle
_{AC_j}\right)  \nonumber \\
&&+\frac 12\left| \Psi ^{+}\right\rangle _{XS}\left( h_j\left| \varphi
_1^j\right\rangle _{AC_j}\mp t_j\left| \varphi _0^j\right\rangle
_{AC_j}\right)  \nonumber \\
&&\pm \frac 12\left| \Phi ^{-}\right\rangle _{XS}\left( t_j\left| \varphi
_1^j\right\rangle _{AC_j}\pm h_j\left| \varphi _0^j\right\rangle
_{AC_j}\right)  \nonumber \\
&&\pm \frac 12\left| \Phi ^{+}\right\rangle _{XS}\left( t_j\left| \varphi
_1^j\right\rangle _{AC_j}\mp h_j\left| \varphi _0^j\right\rangle
_{AC_j}\right) ,  \nonumber
\end{eqnarray}
where $\left| \Psi ^{\pm }\right\rangle _{XS}=\frac 1{\sqrt{2}}\left( \left|
01\right\rangle _{XS}\pm \left| 10\right\rangle _{XS}\right) $, $\left| \Phi
^{\pm }\right\rangle _{XS}=\frac 1{\sqrt{2}}\left( \left| 00\right\rangle
_{XS}\pm \left| 11\right\rangle _{XS}\right) $ are the Bell basis of the
two-qubit system $X\otimes S$. The telecloning process can now be
accomplished by the following procedure.

$\left( i\right) $ Alice performs a Bell-basis measurement of qubits $X$ and 
$S$, obtaining one of the four results $\left| \Psi ^{\pm }\right\rangle
_{XS}$, $\left| \Phi ^{\pm }\right\rangle _{XS}$.

$\left( ii\right) $ Alice use different strategies according to different
measurement results. If the result is $\left| \Psi ^{-}\right\rangle _{XS}$,
the subsystem $AC_j$ is projected precisely into the state $h_j\left|
\varphi _1^j\right\rangle _{AC_j}\pm t_j\left| \varphi _0^j\right\rangle
_{AC_j}=\left| \phi _{\pm }\left( \theta _{N-j}\right) \right\rangle
_A\left| \phi _{\pm }\left( \theta _1\right) \right\rangle _{C_j}$. If $%
\left| \Psi ^{+}\right\rangle _{XS}$ is obtained, $\sigma _z\otimes \sigma
_z $ must be performed on system $AC_j$ since $\left| \varphi
_0^j\right\rangle _{AC_j}$ and $\left| \varphi _1^j\right\rangle _{AC_j}$
obey the following simple symmetry: 
\begin{equation}
\sigma _z\otimes \sigma _z\left| \varphi _i^j\right\rangle _{AC_j}=\left(
-1\right) ^{i+1}\left| \varphi _i^j\right\rangle _{AC_j}.  \eqnum{3.7}
\end{equation}
With above operations, the states of system $AC_j$ are transferred to $%
\left| \phi _{\pm }\left( \theta _{N-j}\right) \right\rangle _A\left| \phi
_{\pm }\left( \theta _1\right) \right\rangle _{C_j}$, just as operation $%
D_j\left( \theta _{N-j},\theta _1\right) $ functions.

$\left( iii\right) $ In the case one of the other two Bell states $\left|
\Phi ^{\pm }\right\rangle _{XP}$ is obtained, the corresponding states are
entangled states. For example, if measurement result is $\left| \Phi
^{-}\right\rangle _{XP}$, the remained states can be written as $\left|
\alpha _{\pm }\right\rangle =\frac{\pm 1}{\sin 2\theta _{N-j+1}}\left(
\left| \phi _{\pm }\left( \theta _{N-j}\right) \right\rangle \left| \phi
_{\pm }\left( \theta _1\right) \right\rangle -\right. $ $\left. \cos 2\theta
_{N-j+1}\left| \phi _{\mp }\left( \theta _{N-j}\right) \right\rangle \left|
\phi _{\mp }\left( \theta _1\right) \right\rangle \right) $, which lie in
the subspace spanned by states $\left\{ \left| \phi _{+}\left( \theta
_{N-j}\right) \right\rangle \left| \phi _{+}\left( \theta _1\right)
\right\rangle \right. $, $\left. \left| \phi _{-}\left( \theta _{N-j}\right)
\right\rangle \left| \phi _{-}\left( \theta _1\right) \right\rangle \right\} 
$. The inner-products show that $\left| \alpha _{\pm }\right\rangle $ are
orthogonal to $\left| \phi _{\mp }\left( \theta _{N-j}\right) \right\rangle
\left| \phi _{\mp }\left( \theta _1\right) \right\rangle $. So they are
entangled states unless $\left| \phi _{+}\left( \theta _{N-j}\right)
\right\rangle \left| \phi _{+}\left( \theta _1\right) \right\rangle $ are
orthogonal to $\left| \phi _{-}\left( \theta _{N-j}\right) \right\rangle
\left| \phi _{-}\left( \theta _1\right) \right\rangle $, which means $\left|
\phi _{\pm }\left( \theta \right) \right\rangle $ are orthogonal. When $%
\left| \phi _{\pm }\left( \theta _1\right) \right\rangle $ are not
orthogonal, Alice and $C_j$ must disentangle the states to the needed states 
$\left| \phi _{\pm }\left( \theta _{N-j}\right) \right\rangle \left| \phi
_{\pm }\left( \theta _1\right) \right\rangle $ simultaneously using only
local operations and classical communication (LQCC). Unfortunately, this
process cannot be deterministic although both transformation $\left| \alpha
_{+}\right\rangle \rightarrow \left| \phi _{+}\left( \theta _{N-j}\right)
\right\rangle \left| \phi _{+}\left( \theta _1\right) \right\rangle $ and $%
\left| \alpha _{-}\right\rangle \rightarrow \left| \phi _{-}\left( \theta
_{N-j}\right) \right\rangle \left| \phi _{-}\left( \theta _1\right)
\right\rangle $ can be deterministically executed according to Nielsen
theorem \cite{Nie99}. In fact, suppose there exists a process $H$ to
accomplish so using only LQCC, the evolution equation of the composite
system of particles $A,C_j$ and the local auxiliary particles $G^A$, $G^{C_j}
$ can be expressed as 
\begin{eqnarray}
&&H\left| \alpha _{\pm }\right\rangle \left| G_0^A\right\rangle \left|
G_0^{C_j}\right\rangle   \eqnum{3.8} \\
&=&\sum_{i=1}^h\sum_{k=1}^l\sqrt{\eta _{ik}}\left| \phi _{\pm }\left( \theta
_{N-j}\right) \right\rangle \left| \phi _{\pm }\left( \theta _1\right)
\right\rangle \left| G_i^A\right\rangle \left| G_k^{C_j}\right\rangle . 
\nonumber
\end{eqnarray}
$H$ is a linear operation, thus we get 
\begin{eqnarray}
&&H\left| \phi _{\pm }\left( \theta _{N-j}\right) \right\rangle \left| \phi
_{\pm }\left( \theta _1\right) \right\rangle \left| G_0^A\right\rangle
\left| G_0^{C_j}\right\rangle   \eqnum{3.9} \\
&=&\left| \alpha _{\pm }\right\rangle \sum_{i=1}^h\sum_{k=1}^l\sqrt{\eta
_{ik}}\left| G_i^A\right\rangle \left| G_k^{C_j}\right\rangle .  \nonumber
\end{eqnarray}
Operation $H$ use only local operations and classical communications which
cannot enhance the entanglement. Obviously no entanglement exists in the
left side of Eq. (3.9), but the right side is an entangled state between
particle $A$, $C_j$. Thus such process $H$ does not exist. However, consider
current experiment technology, only two Bell basis $\left| \Psi ^{\pm
}\right\rangle $ of the four can be identified by interferometric schemes,
with the others $\left| \Phi ^{\pm }\right\rangle $ giving the same
detection signal \cite{Bra95}, so we only need to consider $\left| \Psi
^{\pm }\right\rangle $ in our protocol.

After Alice obtain the state $\left| \phi _{\pm }\left( \theta _{N-j}\right)
\right\rangle _A$, she take it as the input states $\left| \phi _{\pm
}\left( \theta _{N-j}\right) \right\rangle _X$ and use another
three-particle entangled state $\left| \psi ^{j+1}\right\rangle $ to obtain
the states $\left| \phi _{\pm }\left( \theta _{N-\left( j+1\right) }\right)
\right\rangle _A\left| \phi _{\pm }\left( \theta _1\right) \right\rangle
_{C_{j+1}}$ between Alice and $C_{j+1}$, etc,. In the last process, if Alice
wishes to transmit the copies to the associates $C_{N-1}$ and $C_N$, the
system $A$ should be on the side $C_N$. With the series transformations, the
associates $C_1$, $C_2$,..., $C_N$ obtain the states $\left| \phi _{\pm
}\left( \theta _1\right) \right\rangle _{C_j}$ respectively and they finish
the telecloning process.

In the following, we show how to prepare the three-particle entangled state $%
\left| \psi ^j\right\rangle $ represented in Eq. (3.5) by LQCC using GHZ
state as resource. Consider Alice and $C_j$ initially share a GHZ state $%
\left| \xi \right\rangle _{SAC_j}=\frac 1{\sqrt{2}}\left( \left|
000\right\rangle +\left| 111\right\rangle \right) $, to implement the
telecloning process, they must transfer it to the suitable state using only
LQCC. First a local unitary operation $R_y^S\left( \pi /2\right) \otimes
R_y^A\left( -\pi /2\right) \otimes R_y^{C_j}\left( -\pi /2\right) $ is
performed to transfer $\left| \xi \right\rangle _{SAC_j}$ to $\left| \xi
^{^{\prime }}\right\rangle _{SAC_j}=\frac 14\left( \left( \left|
0\right\rangle -\left| 1\right\rangle \right) _S\left( \left| 1\right\rangle
+\left| 0\right\rangle \right) _{AC_j}^{\otimes 2}+\left( \left|
0\right\rangle +\left| 1\right\rangle \right) _S\left( \left| 1\right\rangle
-\left| 0\right\rangle \right) _{AC_j}^{\otimes 2}\right) $. To obtain
required states, local generalized measurement (POVM) is needed, which is
described by operators $M_m$ on corresponding system, satisfying the
completeness relation $\sum_mM_m^{\dagger }M_m=I$. After the measurement,
the results (classical communication) are sent to other system, who performs
a local quantum operation $\varepsilon _m$ on its system according to the
requirement of the transformation task. The operation $\varepsilon _m$ is
conditional on the result $m$ and may be non-unitary.

However, it is difficult to perform the operation $\varepsilon _m$ according
to classical communication in experiment. In the following, we introduce a
method to prepare the initial state by systems $S$, $A$ and $C_j$ performing
local operations respectively without classical communication. In our
protocol, there are two possible final states and both of them can be used
for telecloning with same Bell states $\left| \Psi ^{\pm }\right\rangle $
measured. Define operations $M_{jim}$ $\left( i=1,2,3,m=0,1\right) $ on $S$, 
$A$, and $C_j$ system with matrix representations $M_{j10}=\left( 
\begin{array}{cc}
\sin \theta _{N-j+1} & 0 \\ 
0 & \cos \theta _{N-j+1}
\end{array}
\right) $, $M_{j11}=\left( 
\begin{array}{cc}
\cos \theta _{N-j+1} & 0 \\ 
0 & \sin \theta _{N-j+1}
\end{array}
\right) $, $M_{j20}=\left( 
\begin{array}{cc}
\sin \theta _{N-j} & 0 \\ 
0 & \cos \theta _{N-j}
\end{array}
\right) $, $M_{j21}=\left( 
\begin{array}{cc}
\cos \theta _{N-j} & 0 \\ 
0 & \sin \theta _{N-j}
\end{array}
\right) $, $M_{j30}=\left( 
\begin{array}{cc}
\sin \theta _1 & 0 \\ 
0 & \cos \theta _1
\end{array}
\right) $, $M_{j31}=\left( 
\begin{array}{cc}
\cos \theta _1 & 0 \\ 
0 & \sin \theta _1
\end{array}
\right) $ on the basis $\left| 0\right\rangle $, $\left| 1\right\rangle $
respectively. Note that $M_{ji0}^{\dagger }M_{ji0}+M_{ji1}^{\dagger
}M_{ji1}=I$, therefore those define a generalized measurement on each
system, which may be implemented using standard techniques involving only
projective measurements and unitary transforms \cite{Sch96}. If we consider
a probe $P$ to assist the generalized measurement $M_0=\left( 
\begin{array}{cc}
\sin \theta & 0 \\ 
0 & \cos \theta
\end{array}
\right) $, $M_1=\left( 
\begin{array}{cc}
\cos \theta & 0 \\ 
0 & \sin \theta
\end{array}
\right) $, the unitary operator acting on the particle and the probe can be
represented as $\left( 
\begin{array}{cc}
R_y\left( -\pi +2\theta \right) & 0 \\ 
0 & R_y\left( -2\theta \right)
\end{array}
\right) $ on the basis $\left\{ \left| 0P_0\right\rangle ,\left|
0P_1\right\rangle ,\left| 1P_0\right\rangle ,\left| 1P_1\right\rangle
\right\} $, where $R_y\left( \theta \right) =\left( 
\begin{array}{cc}
\cos \frac \theta 2 & \sin \frac \theta 2 \\ 
-\sin \frac \theta 2 & \cos \frac \theta 2
\end{array}
\right) $ is a rotation by $\theta $ around $\hat{y}$. If the measurement
result gives $m=1$ for a system, then a rotation $\sigma _x$ is performed on
this system. Let $\left| \xi _{\left( -1\right) ^{k+p+t}}\right\rangle
_{SAC_j}$ denote the state after the measurement and local $\sigma _x$,
given that outcome $k$, $p$, $t$ occurred for $A$, $C_j$, $S$ system
respectively, then%
%TCIMACRO{\TeXButton{widetext}{\widetext} }
%BeginExpansion
\widetext%
%EndExpansion
\begin{equation}
\left| \xi _{\left( -1\right) ^{k+p+t}}\right\rangle _{SAC_j}=\left\{ 
\begin{array}{c}
\frac 1{\sqrt{2}}\left( \left| 0\right\rangle _S\left| \varphi
_1^j\right\rangle _{AC_j}-\left| 1\right\rangle _S\left| \varphi
_0^j\right\rangle _{AC_j}\right) \text{, when }\left( -1\right) ^{k+p+t}=1
\\ 
\kappa \left( \frac{t_j}{h_j}\left| 0\right\rangle _S\left| \varphi
_0^j\right\rangle _{AC_j}-\frac{h_j}{t_j}\left| 1\right\rangle _S\left|
\varphi _1^j\right\rangle _{AC_j}\right) \text{, when }\left( -1\right)
^{k+p+t}=-1
\end{array}
\right.  \eqnum{3.10}
\end{equation}
%TCIMACRO{\TeXButton{narrowtext}{\narrowtext}}
%BeginExpansion
\narrowtext%
%EndExpansion
where $\kappa =\sqrt{\frac{1-\cos ^22\theta _{N-j+1}}{2(1+\cos ^22\theta
_{N-j+1})}}$. The probability to obtain the first state $\left| \xi
_1\right\rangle _{SAC_j}$ is $p_1=\frac{\sin ^22\theta _{N-j+1}}2$ and the
second $\left| \xi _{-1}\right\rangle _{SAC_j}$ is $p_{-1}=\frac{1+\cos
^22\theta _{N-j+1}}2$. The first state in Eq. (3.10) is exactly the state in
Eq. (3.5) and the second state can also be used for telecloning. In fact,
the combined states of systems $XSAC_j$ can be rewritten in a form that
singles out the Bell basis of qubit $X$ and $S$ as 
\begin{eqnarray}
&&\left| \psi ^{\pm j}\right\rangle _{XSAC_j}^{^{\prime }}  \eqnum{3.11} \\
&=&\mp \frac \kappa {\sqrt{2}}\left| \Psi ^{-}\right\rangle _{XS}\left(
h_j\left| \varphi _1^j\right\rangle _{AC_j}\pm t_j\left| \varphi
_0^j\right\rangle _{AC_j}\right)  \nonumber \\
&&\pm \frac \kappa {\sqrt{2}}\left| \Psi ^{+}\right\rangle _{XS}\left(
h_j\left| \varphi _1^j\right\rangle _{AC_j}\mp t_j\left| \varphi
_0^j\right\rangle _{AC_j}\right)  \nonumber \\
&&+\frac \eta {\sqrt{2}}\left| \Phi ^{-}\right\rangle _{XS}\left(
h_j^3\left| \varphi _1^j\right\rangle _{AC_j}\pm t_j^3\left| \varphi
_0^j\right\rangle _{AC_j}\right)  \nonumber \\
&&-\frac \eta {\sqrt{2}}\left| \Phi ^{+}\right\rangle _{XS}\left(
h_j^3\left| \varphi _1^j\right\rangle _{AC_j}\mp t_j^3\left| \varphi
_0^j\right\rangle _{AC_j}\right)  \nonumber
\end{eqnarray}
where $\eta =\frac{2\kappa }{\sin 2\theta _{N-j+1}}$. Obviously the first
two terms can be transferred to the target states using same unitary
operations as those in Eq. (3.6) and states $h_j^3\left| \phi
_0^j\right\rangle _{AC_j}\pm t_j^3\left| \phi _1^j\right\rangle
_{AC_j}=\left| \phi _{\pm }\left( \theta _{N-j}\right) \right\rangle \left|
\phi _{\pm }\left( \theta _1\right) \right\rangle +\cos 2\theta
_{N-j+1}\left| \phi _{\mp }\left( \theta _{N-j}\right) \right\rangle \left|
\phi _{\mp }\left( \theta _1\right) \right\rangle $ need not be considered.

The probabilistic quantum cloning process via GHZ states is illustrated in
Fig. 3(A) and Fig. 3(B) for the case $M=1$, $N=2$. 
\[
\text{Fig. 3(A) and Fig. 3(B)} 
\]

The unitary-reduction operation $U$ in Eq. (3.2) and the generalized
measurements $M_{jim}$ can be implemented using linear optical components,
i.e., polarizing beam splitter (PBS) and polarization rotation (PR). In Ref. 
\cite{Cerf98}, Cerf {\it et al.} constructed the location controlling
polarization (LCP) NOT gate using a PR. A general LCP unitary rotation can
also be executed similarly. The polarization controlling location (PCL) NOT\
gate is performed by the use of a PBS. However, a PCL unitary rotation need
two PBS and some PR since direct rotation of location qubit is impossible.
Generally, a PCL unitary rotation can be represented as $V=\left( 
\begin{array}{cc}
R_y\left( \xi \right) & 0 \\ 
0 & R_y\left( \chi \right)
\end{array}
\right) $ on the orthogonal basis $\left\{ \left| 0\right\rangle \left|
P_0\right\rangle ,\left| 0\right\rangle \left| P_1\right\rangle ,\left|
1\right\rangle \left| P_0\right\rangle ,\left| 1\right\rangle \left|
P_1\right\rangle \right\} $, with $\left| 0\right\rangle ,\left|
1\right\rangle $ denoted as the polarization qubit and $\left|
P_0\right\rangle ,\left| P_1\right\rangle $ as the location qubit. $V$ can
be decomposed into $V=V_1V_2V_3V_2V_1$, where $V_1$ is a LCP-NOT gate, $V_2$
is a PCL-NOT gate and $V_3$ represents a LCP unitary operation that performs 
$R_y\left( \xi \right) $ on the polarization qubit if the location qubit is
on $\left| P_0\right\rangle $, and $R_y\left( -\chi \right) $ if the
location qubit on $\left| P_1\right\rangle $. So operation $V$ can be
implemented using linear optical components as that in Fig. 4. 
\[
\text{Fig. 4} 
\]

Each generalized measurement $M$ gives two output paths $0$ and $1$ and
eight possible results may be output for the three photons while they only
represent two possible final states $\left| \xi _1\right\rangle _{SAC_j}$
and $\left| \xi _{-1}\right\rangle _{SAC_j}$. By the use of fiber the two
paths for each $M$ can be convert into one, which means tracing out over the
location qubit, and the final state of the three photons turns into the
mixed state $\rho _{SAC_j}=p_1\left| \xi _1\right\rangle \left\langle \xi
_1\right| +p_{-1}\left| \xi _{-1}\right\rangle \left\langle \xi _{-1}\right| 
$. However, after Bell basis measurement of the tensor product state $\left|
\phi _{\pm }\left( \theta _{N-j+1}\right) \right\rangle _X\left\langle \phi
_{\pm }\left( \theta _{N-j+1}\right) \right| \otimes \rho _{SAC_j}$, the
final states are still $h_j\left| \varphi _1^j\right\rangle _{AC_j}\pm
t_j\left| \varphi _0^j\right\rangle _{AC_j}$ and $h_j\left| \varphi
_1^j\right\rangle _{AC_j}\mp t_j\left| \varphi _0^j\right\rangle _{AC_j}$
corresponding to $\left| \Psi ^{-}\right\rangle _{XS}$ and $\left| \Psi
^{+}\right\rangle _{XS}$ because of Eq. (3.6) and (3.11).

Let us compare the efficiency of above telecloning process and that using
Tele-C-NOT gates \cite{Got99}. To complete a Tele-C-NOT operation, two GHZ
states and three Bell basis measurement are need, which yields $1/8$
probability. Performing a $D_j\left( \theta _{N-j},\theta _1\right) $
operation needs three C-NOT gates \cite{Che99,Zhang00}, that is, Alice only
has probability of $\frac 1{512}$ to succeed. While our protocol use one GHZ
states and yields the probability 
\begin{eqnarray}
p &=&p_1\times \frac 12+p_{-1}\times \kappa ^2  \eqnum{3.12} \\
&=&\frac{\sin ^22\theta _{N-j+1}}2=\frac{1-\cos ^{2\left( N-j+1\right)
}2\theta }2.  \nonumber
\end{eqnarray}
When $\theta $ is not too small, the success probability is not too low. If
we do not consider the preparation of three-particle entanglement states,
the efficiency of Tele-$D_j\left( \theta _{N-j},\theta _1\right) $ is $50\%$%
, which is exactly the efficiency of Bell measurement. If we have enough GHZ
states, we can prepare enough required three-particle entangled states. In
the initial information compress process, we adopt the $M\times \left(
1\rightarrow N\right) $ cloning strategy. Using this strategy, more than one 
$\left| \phi _{\pm }\left( \theta _N\right) \right\rangle $ can be obtained.
So if the Tele-$D_j\left( \theta _{N-j},\theta _1\right) $ operation fails
to one $\left| \phi _{\pm }\left( \theta _N\right) \right\rangle $, we have
chance to use another and that increases the success probability. The
overall cloning probability of our protocol (not include that in states
preparation) can be represented as 
\begin{equation}
P=\sum_{k=1}^MC_M^k\gamma _{1N}^k\left( 1-\gamma _{1N}\right) ^{M-k}\left(
1-\left( 1-\left( \frac 12\right) ^{N-1}\right) ^k\right) .  \eqnum{3.13}
\end{equation}
$P\ $decreases with the increase of $N$, therefore we often adopt $%
1\rightarrow 2$ cloning strategy in practice.

Up to this point, our discussion has assumed that the initially shared
three-partite entangled states are pure GHZ states. Suppose, however, that $%
\left| \xi \right\rangle _{SAC_j}$ is corrupted a little by decoherence
before it is made available to the systems $S$, $A$ and $C_j$, so they
receive a density matrix $\sigma $ instead. What can we say about the final
states and the probabilities of success? We argue that the final states and
the probabilities do not change too much if the windages of initial states
are not too large.

We discuss this problem using the {\it trace distance}, a metric on
Hermitian operators defined by $T(A,B)\equiv \mbox{Tr}(|A-B|)$, where $|X|$
denotes the positive square root of the Hermitian matrix $X^2$. The trace
distance is a quantity with a well-defined {\it operational meaning} as the
probability of making an error distinguishing two states \cite{Fuch99}. In
this sense it may reflect the possible physical approximation between the
states: the value of the trace distance smaller, the two states more
similar. A direct example is that for pure states $\psi $ and $\phi $ the
trace distance and the fidelity are related by a simple formula, 
\begin{eqnarray}
T(\psi ,\phi )=2\sqrt{1-F(\psi ,\phi )}.  \eqnum{3.14}
\end{eqnarray}

Ruskai \cite{Rus94} has shown that the trace distance contracts under
physical processes. More precisely, if $\varpi $ and $\sigma $ are any two
density operators, and if $\varpi ^{\prime }\equiv {\cal E}(\varpi )$ and $%
\sigma ^{\prime }\equiv {\cal E}(\sigma )$ denote states after some physical
process represented by the (trace-preserving) quantum operation ${\cal E}$
occurs, then 
\begin{eqnarray}
T(\varpi ^{\prime },\sigma ^{\prime })\leq T(\varpi ,\sigma ).  \eqnum{3.15}
\end{eqnarray}

So, after the telecloning process, the change of the final states is limited
by the trace distance between initial states $\left| \xi \right\rangle
_{SAC_j}\left\langle \xi \right| $ and $\sigma $, and the continuity of
probability also promises the less alteration of the successful
probabilities represented by Eq. (3.12) and Eq. (3.13). Of course, the final
states may not be the pure cloning states we required at this situation. It
may be a mixed states resembling the cloning states with the accuracy
dependent on the windage of the initial states.

Such telecloning process can also be accomplished using a multiparticle
entangled state, similar as that has been shown in \cite{Mur99}. The quality
of our method is that only three-particle entanglement is used. In this
scheme, we use local generalized measurements and Bell basis measurement to
avoid the interactions between particles, so it may be feasible in current
experiment condition.

\section{Summary}

In summary, we have presented a probabilistic quantum cloning scheme using
GHZ states, Bell basis measurements, single-qubit unitary operations and
generalized measurements, all of which are within the reach of current
technology. We considered different strategies and propose the concept of 
{\it Probability Spectrum} to describe them. For two most important, we show
that $M$ entries $1\rightarrow N$ cloning process give more copies than one $%
M\rightarrow N$ process at the price of higher probability of failure.
Compared to another possible scheme via Tele-C-NOT \cite{Got99} gate, our
scheme may be feasible in experiment to {\it clone the states of one
particle to those of two different particles} with higher probability and
less GHZ resource.

\begin{center}
{\bf ACKNOWLEDGMENT}
\end{center}

This work was supported by the National Natural Science Foundation of China.

\begin{center}
{\bf Figure Captions:}
\end{center}

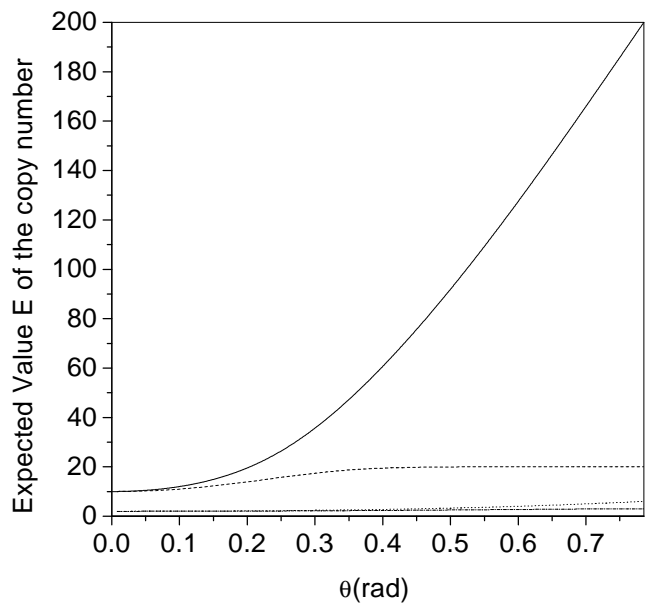
Fig. 1: The expected values of copy number for the two different strategies.
Angle $\theta $ is corresponding to initial states set $\left\{ \cos \theta
\left| 1\right\rangle \pm \sin \theta \left| 0\right\rangle \right\} $. Here
Solid line, Dashed line, Dotted line and Dash-Dotted line denote $10\times
\left( 1\rightarrow 20\right) $, $1\times \left( 10\rightarrow 20\right) $, $%
2\times \left( 1\rightarrow 3\right) $ and $1\times \left( 2\rightarrow
3\right) $ cloning strategies respectively.

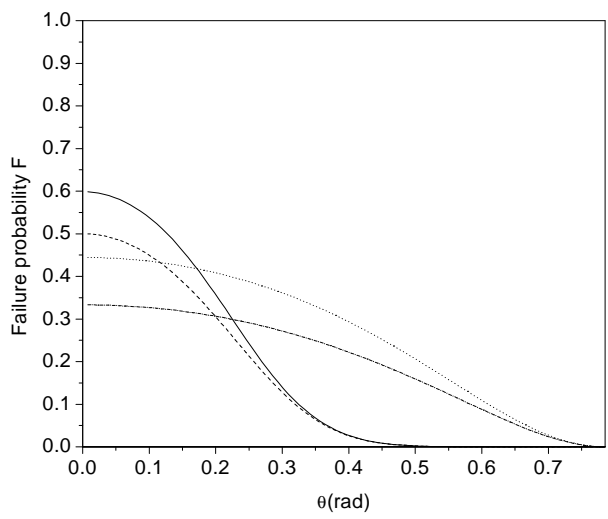
Fig. 2: The failure probabilities for the two different strategies. The four
kinds of lines represent the same strategies as those in Fig. 1.

Fig. 3(A): The logic network of $1\rightarrow 2$ probabilistic cloning via
GHZ state. Alice and her associate $C_1$, $C_2$ initially share a GHZ state
consisting of the qubit $S$ (the port), $C_1$ and $C_2$ (outputs, or `copy
qubits'). Alice successfully transforms the initial states $\cos \theta
\left| 1\right\rangle _X\pm \sin \theta \left| 0\right\rangle _X$ to $\cos
\theta _2\left| 1\right\rangle _X\pm \sin \theta _2\left| 0\right\rangle _X$
if the probe (the location qubit of the photon $X$) results in $\left|
P_0\right\rangle $, where the parameters $\cos ^22\theta _2=\cos 2\theta $, $%
\omega =\arccos \sqrt{\frac{\left( 1+\cos ^22\theta \right) }{\left( 1+\cos
2\theta \right) ^2}}$. Using the unitary rotation $R_y\left( \varsigma
\right) $ and generalized measurement $M\left( \theta \right) $, Alice and $%
C_1$, $C_2$ transform GHZ state to the required three-particle entangled
state in the form Eq. (3.10). Then Alice performs a Bell measurement of the
port $S$ along with `input' qubit $X$ and has $25\%$ probability to obtain $%
\left| \Psi ^{-}\right\rangle $ or $\left| \Psi ^{+}\right\rangle $
respectively; subsequently, the receivers $C_1$ and $C_2$ do no operation or 
$\sigma _x$ rotations on the output qubits, obtaining two perfect quantum
clones. The implementation of generalized measurement $M\left( \theta
\right) $ is illustrated in Fig. 3(B).

Fig. 3(B): The implementation of generalized measurement $M\left( \theta
\right) $ in Fig. 3(A). The location qubit of the photon is adopted as the
probe $P$.

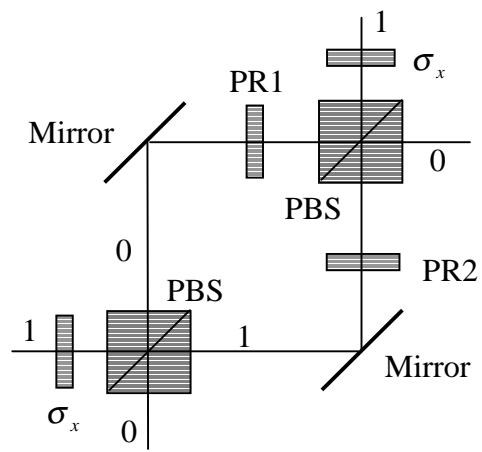
Fig. 4: Optical simulation of PCL unitary rotation by the use of two
polarizing beam splitters and some polarizing rotators, where PR1 performs
operation $R_y\left( \xi \right) $ and PR2 executes operation $R_y\left(
-\chi \right) $.%
%TCIMACRO{
%\TeXButton{widetext}{\widetext
%}}
%BeginExpansion
\widetext
%
%EndExpansion

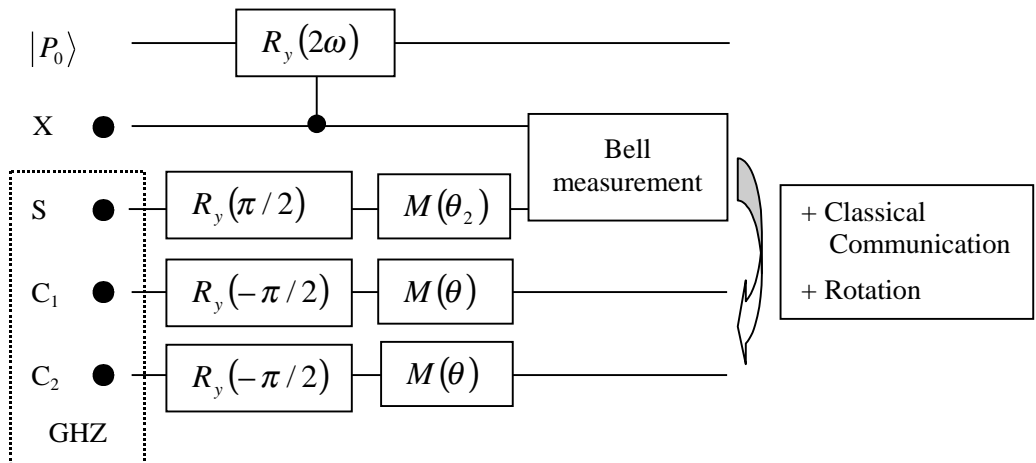
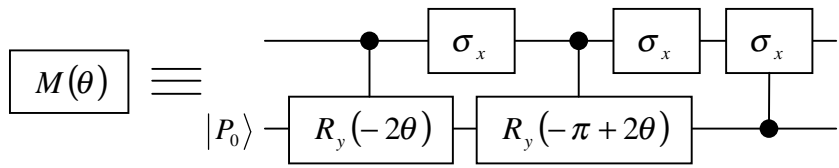
\end{document}